\documentclass[3p,times,procedia]{elsarticle}
\usepackage{nupha_ecrc}

\volume{00}

\firstpage{1}

\journalname{Nuclear Physics A}

\runauth{}

\jid{nupha}

\jnltitlelogo{Nuclear Physics A}

\usepackage{amssymb}

\usepackage[figuresright]{rotating}

\begin{document}

\begin{frontmatter}



\dochead{XXVIIth International Conference on Ultrarelativistic Nucleus-Nucleus Collisions\\ (Quark Matter 2018)}

\title{Neutron stars meet constraints from high and low energy nuclear physics}

\author[1,2]{V. Sagun}
\author[1,3]{I. Lopes}
\author[4,2]{A. Ivanytskyi}
\address[1]{Centro de Astrof\'{\i}sica e Gravita\c c\~ao  - CENTRA,
Departamento de F\'{\i}sica, Instituto Superior T\'ecnico - IST, Universidade de Lisboa - UL, Av. Rovisco Pais 1, 1049-001 Lisboa, Portugal}
\address[2]{Bogolyubov Institute for Theoretical Physics, Metrologichna str. 14$^B$, Kyiv 03680, Ukraine}
\address[3]{Institut d'Astrophysique de Paris, Sorbonne Universit\'e,  98 bis Boulevard Arago, Paris F-75014, France}
\address[4]{Department of Fundamental Physics, University of Salamanca, Plaza de la Merced s/n 37008, Spain} 

\footnote{violettasagun@tecnico.ulisboa.pt, ilidio.lopes@tecnico.ulisboa.pt, oivanytskyi@usal.es}

\begin{abstract}
A novel equation of state used for analysis of the heavy ion collision experimental data is generalized to also describe the matter inside neutron stars. This approach differs from others by including an induced surface tension caused by the short range repulsion between particles accounted by presence of their hard cores. This new equation of state respects thermodynamic consistency and is free from the usual causality problems. This makes it applicable
over a wide range of temperatures and baryon densities, including the ones inside neutron stars. Despite small number of parameters the present equation of state reproduces properties of the saturated nuclear matter and is consistent with the proton flow constraint as well as with astrophysical data obtained from neutron star observations. Accordingly, we found the sets of parameters that agree with the same ones obtained from the heavy ion collision data analysis \footnote{ Two sets of the tabulated EoS are publicly available at the https://centra.tecnico.ulisboa.pt/network/costar/files/ist/}.
\end{abstract}


\end{frontmatter}


\section{Introduction}
\label{Intro}
Neutron stars (NS) are among the most dense objects in the Universe that are directly observable  and which  properties are determined by an interplay of gravitational and strong nuclear forces \cite{Lattimer}. This explains the recent interest in studying NSs, since such stars are important natural laboratories to test both General Relativity and the modern theory of strong interactions. Recently this interest was significantly enhanced by the first observation of the merger of two NSs~\cite{Abbott}. On the other hand, the exploration of the QCD phase diagram within ongoing and planned experimental programs on heavy ion collisions (HIC) \cite{Bugaev1} also requires a detailed information about properties of strongly interacting matter. Experience gained by astrophysical and high energy physics communities shows that unified approaches to study the QCD matter in regions typical for NS and HIC is of principal importance for understanding processes inside compact astrophysical objects and hot hadronic medium. Development of such an unified treatment of matter with strong interaction requires an equation of state (EoS) applicable in a wide range of baryonic densities and temperatures. In this work we briefly report a recent development towards the formulation of such an EoS. 
\vspace*{-.1cm}

\section{Equation of state of dense matter}
\label{Model}

\begin{figure}[!]
\centering
\includegraphics[width=0.43\columnwidth]{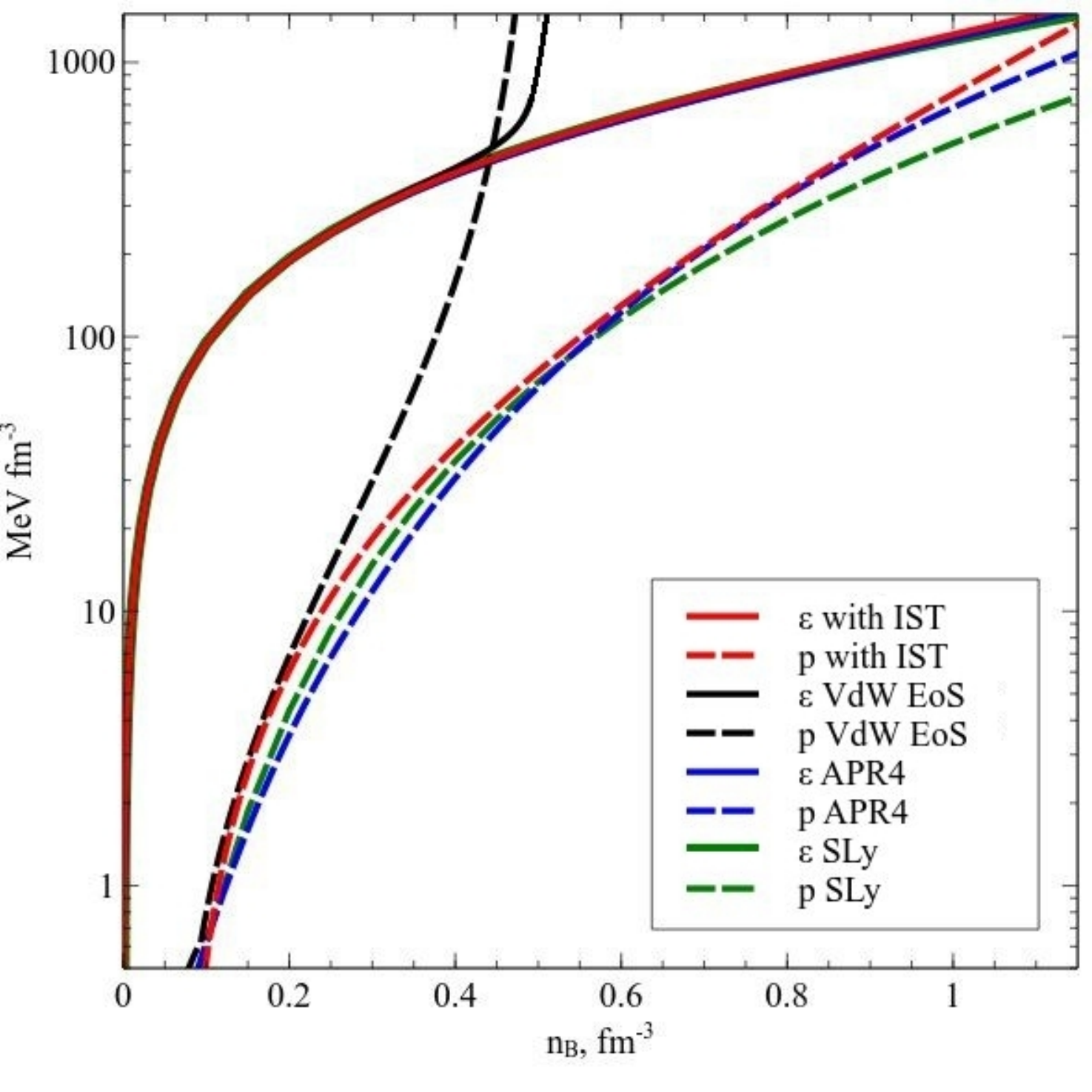}
\hspace*{1.5cm}
\includegraphics[width=0.43\columnwidth]{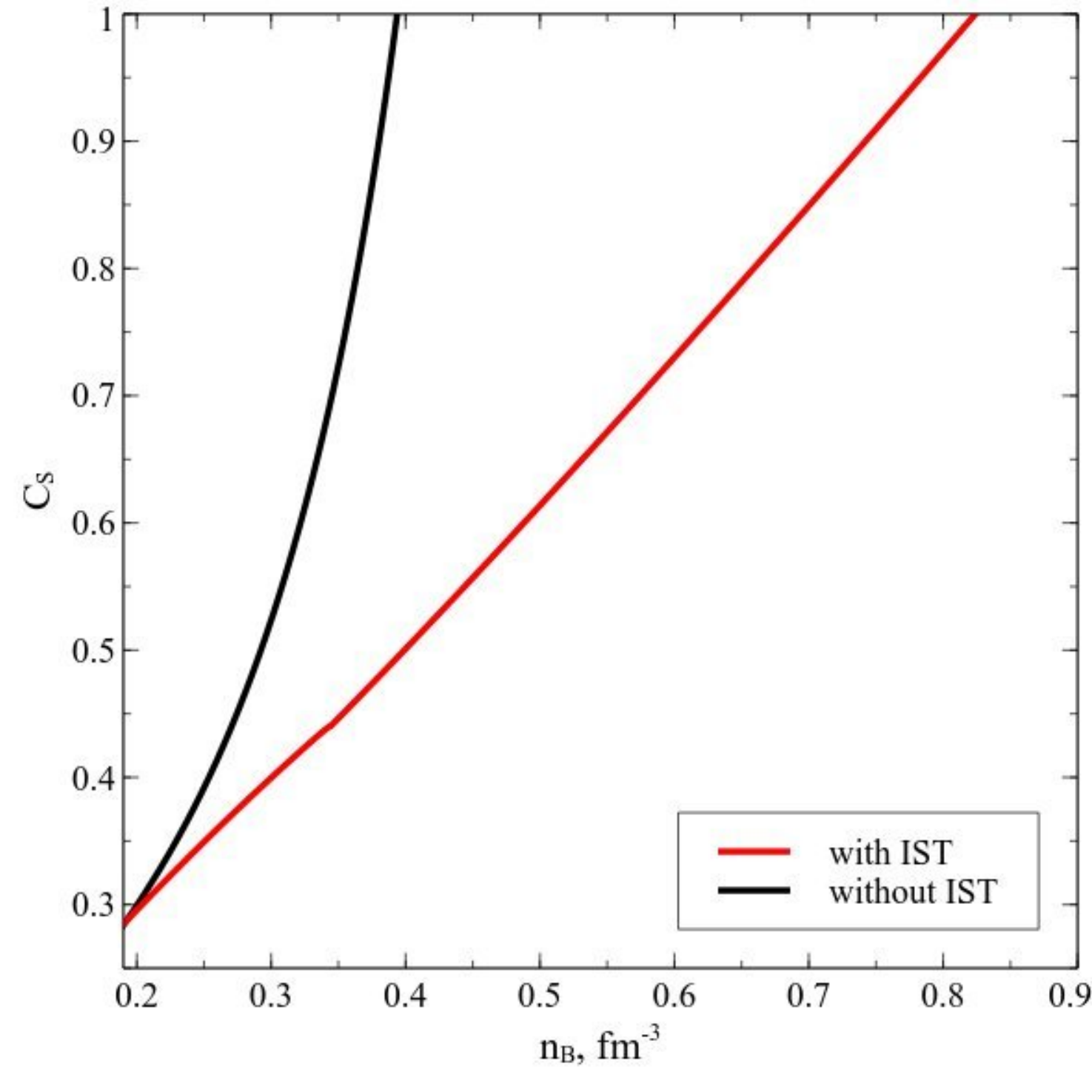}  
\vspace*{-.3cm}  
\caption{{\bf Left panel:} energy density $\epsilon$ (solid curves) and pressure $p$ (dashed curves) as functions of the baryonic density $n_B$. {\bf Right panel:} speed of sound $C_s$ as functions of the baryonic density $n_B$. Calculations are performed for electrically neutral IST EoS (red curves) and EoS with the Van der Waals parametrization of hard core repulsion (black curves) with the parameter set A from Table \ref{tab1}.}
\vspace*{-.2cm}
\label{fig1}
\end{figure}

A delicate balance between interparticle repulsion and attraction is a key element to formulate a realistic EoS of dense baryonic matter. In our approach we account for the long range attraction of the mean field type as well as for the strong short range repulsion between particles. The latter is modelled by the hard core repulsion within the framework of induced surface tension (IST) which was successfully applied for modelling properties of symmetric nuclear matter \cite{Sagun2014}, analysis of hadron yields measured in HIC from AGS \cite{Sagun2017} to ALICE \cite{Bugaev2017} energies and to description of compact astrophysical objects \cite{Sagun2017APJ}. This approach allows us to go beyond the usual Van der Waals approximation and to safely describe baryonic matter at high densities. IST generated by interparticle interaction in thermal medium ensures correct values of the second, third and fourth virial coefficients of hard spheres controlled by a single parameter $\alpha$ \cite{Sagun2014}. It is remarkable that the set up of the model with $\alpha=1.245$ not only reproduces above virial coefficients but also leads to the widest causality range of the hadronic system \cite{Bugaev2017}. This fact gives entirely independent justification to the IST framework. As it is seen from Fig. \ref{fig1}, the IST EoS is significantly softer than the one with the Van der Waals parametrization  of the hard core repulsion. This provides a wider range of the present EoS causality and allows its application for modelling the NS interiors characterized by high baryonic densities.

\begin{figure}[!]
\centering
\includegraphics[width=0.43\columnwidth]{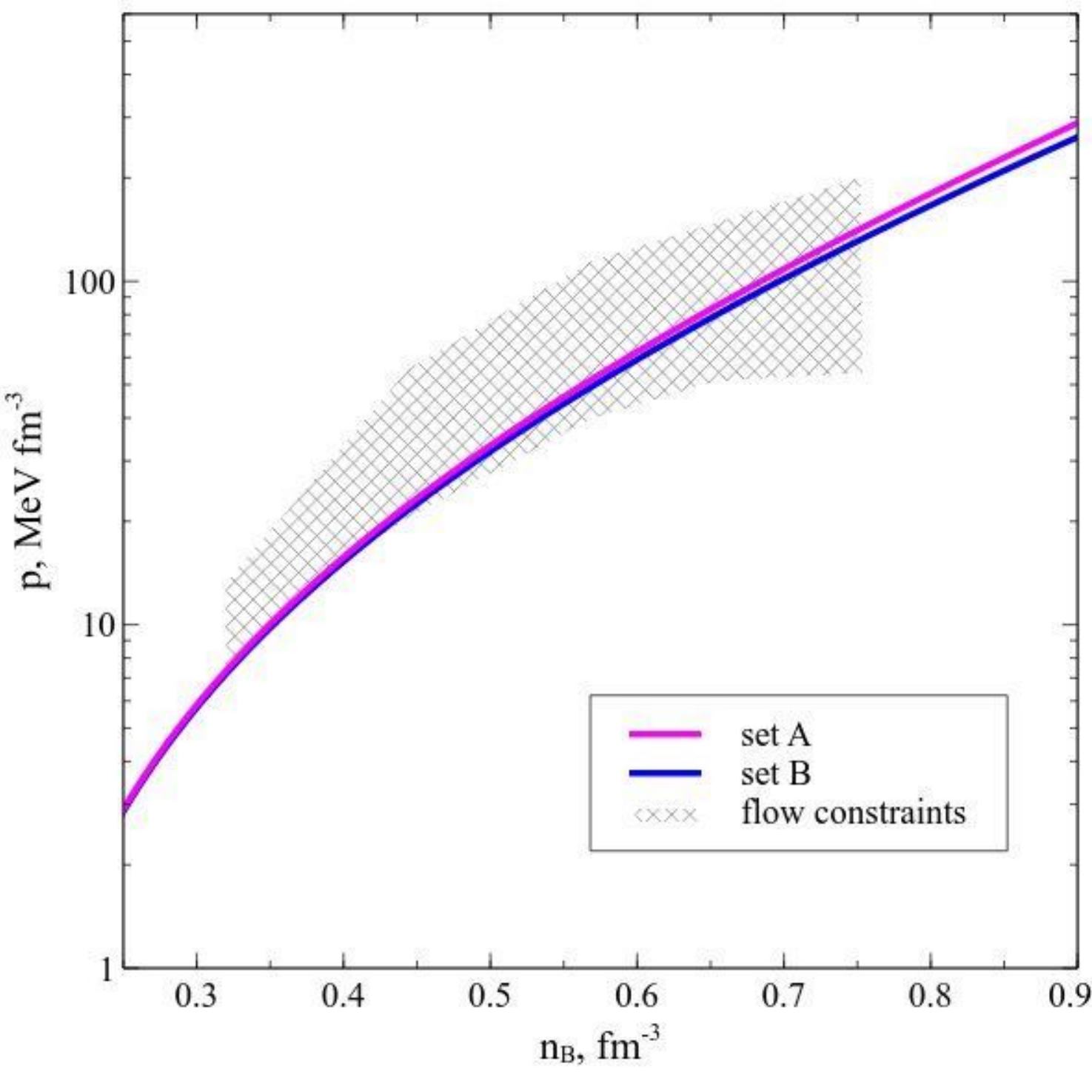}\hspace*{1.5cm}
\includegraphics[width=0.43\columnwidth]{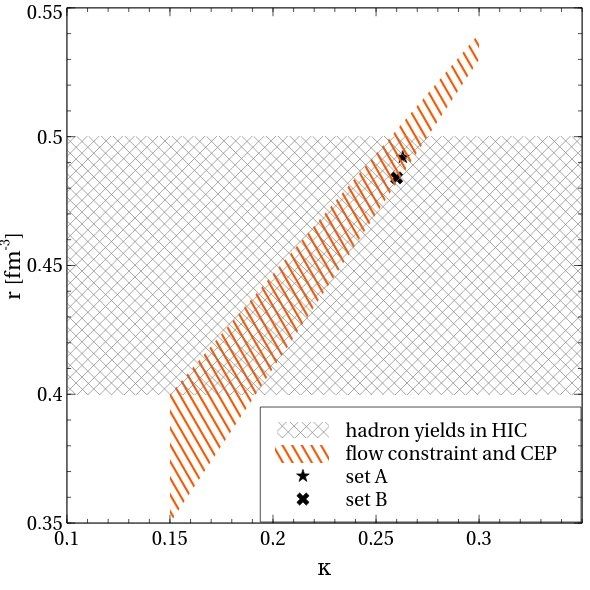} 
\vspace*{-.3cm}
\caption{{\bf Left panel:} pressure $p$ of symmetric nuclear matter as function of baryonic density $n_B$ for the parameter sets from Table \ref{tab1}, the shaded area corresponds to the flow constraint. {\bf Right panel:} restrictions on values of the parameter $\kappa$ and nucleon hard core radius $r$ found from the analysis hadron multiplicities measured in HIC and flow constraint.}\vspace*{-.2cm}
\label{fig2}
\end{figure}

Another important ingredient of our EoS is the mean field attraction which generates a negative correction to the pressure and at small densities is proportional to $n_B^{1+\kappa}$ with $n_B$ being a baryon density, $C_d^2$ a proportionality coefficient and $0<\kappa<1$ \cite{Ivanytskyi2018}. Thermodynamic consistency (see \cite{Rischke} for details) of the given approach is provided by the positive shift of the baryon chemical potential which at small densities is proportional to $n_B^\kappa$. Repulsion of the mean field type is not explicitly introduced to the present EoS because of the fact that even at five nuclear saturation densities the average separation between particles is quite large. At such a regime the microscopic interaction potentials are essentially attractive. This allows one to safely account for the residual repulsive interaction as a hard core correction. Properties of symmetric nuclear matter are highly sensitive to details of the mean field attraction. At ground state characterized by baryonic density $n_0=0.16~fm$ and saturated value of binding energy per nucleon $16~MeV$ repulsion caused by the particle hard cores and Fermi momenta is compensated by this attraction resulting in a zero total pressure. In Ref. \cite{Ivanytskyi2018} it was shown that consistency of the IST EoS with the flow constraint on the symmetric nuclear matter EoS \cite{Danielewicz} implies $0.1<\kappa<0.3$. At the same time, additional constraints on the values of the nuclear incompressibility factor $K_0=200-300~MeV$ and requirement of a realistic critical endpoint of nuclear matter makes this interval even narrower, with $\kappa$ varying from $0.15$ (with $r=0.35-0.4~fm$) to $0.30$ (with $r=0.53-0.54~fm$). These results can be complemented by the conclusions  drawn from the analysis of the HIC data on hadron multiplicities that nucleon hard core radius $r$ lies between $0.4$ and $0.5~fm$ \cite{Sagun2014,Sagun2017,Bugaev2017}. As it is seen from Fig. \ref{fig2}, the resulting region in the $\kappa-r$ plain is rather compact. We found two parameterizations of the IST EoS which generate the maximal NS mass above two solar ones (set A and set B from Table \ref{tab1}). They are fully consistent with the flow constraint (see Fig. \ref{fig2}). It is remarkable that these sets correspond to two points which are located well inside the above region (see Fig. \ref{fig2}). This means that the IST EoS fitted to the high and low energy nuclear physics data is fully consistent with the highest observed mass of NS \cite{Antoniadis}.

\begin{table*}[t]
\caption{\label{tab1} Parameters of the IST EoS}
\begin{center}
{\footnotesize
\begin{tabular}{|c|l|l|l|l|l|l|l|l|l|l|l|}
\hline	
IST EoS & $r$ & $\alpha$ & $\kappa$ & $B^s$& $A^s$ &$C_{d}^{2} $ & $U_0$  & $K_0$  &  $J$ & $L$  & $M_{max}$\\
  &  $fm$  & $-$ & $-$  & $fm^{3}$ & $ MeV \cdot fm^{3}$& $ MeV \cdot fm^{3\kappa}$& MeV & MeV & MeV  & MeV & $M_{\odot}$ \\
\hline
{\bf Set A} & 0.492  & 1.245  & 0.263 & 3.5 & 16.896 & 143.564 & 147.456 &  200.03 & 30.0 & 114.91 & 2.229 \\
\hline
{\bf Set B} & 0.484  & 1.245  & 0.26 & 4.5 & 14.762 & 144.042 & 150.97 &  200.00 & 30.0 & 113.28 & 2.189 \\
\hline
\end{tabular}}\vspace*{-.5cm}
\end{center}
\end{table*}

The NS interiors are characterized by different fractions of neutrons and protons or, equivalently, by non zero nuclear asymmetry $I$. It generates a positive contribution to the pressure and corresponds to additional repulsion. In order to take under control both nuclear symmetry energy $J$ and its slope $L$ at saturation density, we adopted the mean field type parametrization of the asymmetry pressure which at small $n_B$ behaves as $\frac{A^s(In_B)^2}{1+(B^sIn_B)^2}$. Utilization of the scheme developed in Ref. \cite{Rischke}  provided thermodynamic consistency of this EoS.
\vspace*{-.5cm}

\begin{figure}[!]
\centering
\includegraphics[width=0.43\columnwidth]{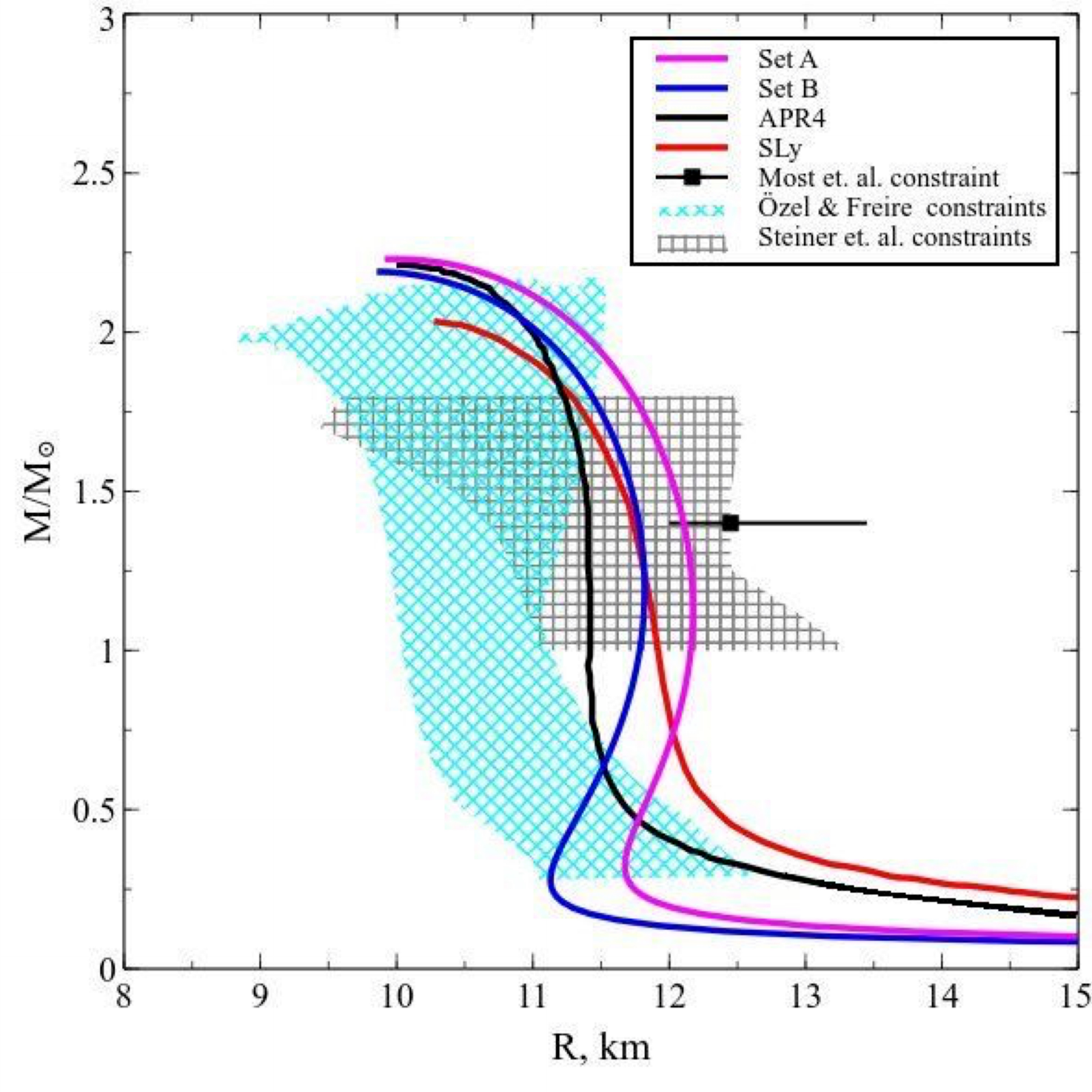}
\hspace*{1.5cm}
\includegraphics[width=0.43\columnwidth]{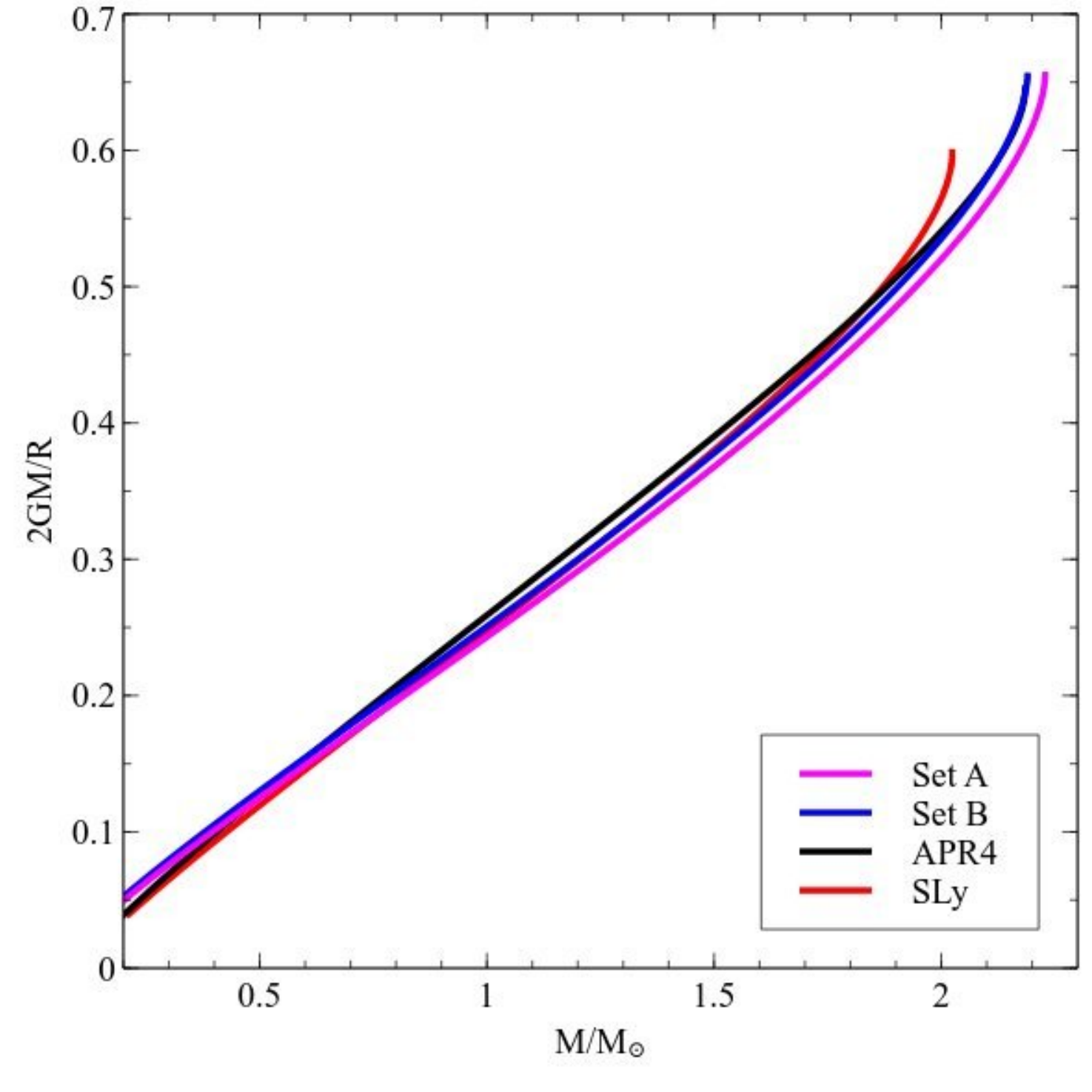} 
\vspace*{-.3cm} 
\caption{{\bf Left panel:} gravitational mass $M$ of NS as function of its radius $R$. {\bf Right panel:} compactness of NS $2GM/R$ as function of its mass $M$. Calculations are performed for sets A and B of the IST EoS (see Table \ref{tab1}). The set A provides the best description of the astrophysical constraints, while the set B corresponds to the mass-radius relation close to the ones of the APR4 and SLy EoSs.}
\vspace*{-.2cm}
\label{fig3}
\end{figure}

\section{Description of neutron stars}
\label{NS}

Masses and corresponding radii of NS were found by numerical integration of the Tolman - Oppenheimer - Volkoff equation \cite{Tolman,Oppenheimer}. For this purpose EoS of dense nuclear matter was supplemented by the condition of electric neutrality provided by the presence of electrons. The latter were treated as non interacting massless particles existing in $\beta$-equilibrium with the surrounding medium. In order to mimic the structure of the atomic crust on the top of the NS's core, a polytropic EoS with $\gamma=\frac{4}{3}$ was applied at very small densities. Such an envelope affects the NS radius and mass only when the latter is below $0.2M_\odot$. Found mass-radius diagram is shown on Fig. \ref{fig3}. Set A of the model parameters allows the IST EoS to fit astrophysical constraints \cite{Most,Ozel,Steiner} in the best way. At the same time, set B generates a mass-radius diagram close to the ones of the APR4 \cite{Akmal} and SLy \cite{Douchin} EoSs, which are widely used at present. Despite valuable differences in the mass-radius diagrams of different EoSs, their compactnesses shown on Fig. \ref{fig3} are similar. Small differences appear only at high masses of NS. It is interesting because such a regime corresponds to the highest densities of the NS core, when effects of the IST become the most important.
\vspace*{-.1cm}
\section{Conclusions}
\label{Concl}
Based on this phenomenological EoS that is able to simultaneously satisfy various constraints at  high and low energy physics, we studied the mass-radius relation and the compactness of NSs.  Special attention was paid to account for the short-range repulsion between nucleons within the IST framework. This has allowed us to extend the applicability range of this EoS up to the densities typical for the NS interior. We found two sets of the model parameters, which on the one hand are consistent with properties of normal nuclear matter, flow constraint and HIC data on hadron yields and, on the other hand, allows present EoS to fit existing astrophysical data. We do hope that the present approach will be successfully used to investigate the strongly interacting matter in a wide range of densities and temperatures.

\section{Acknowledgments}
\label{Acknow}
VS and IL thanks the Funda\c c\~ao para a Ci\^encia e Tecnologia (FCT), Portugal, for the financial support to the
Centro de Astrof\'{\i}sica e Gravita\c c\~ao (CENTRA), Instituto Superior T\'ecnico, Universidade de Lisboa
through the Grant No. UID/FIS/00099/2013. The work of AI was performed within the project SA083P17 of Universidad de Salamanca launched by the Regional Government of Castilla y Leon and the European Regional Development Fund. AI and VS also acknowledge a partial support by the Program of Fundamental Research in High Energy and Nuclear Physics launched by the Section of Nuclear Physics of the National Academy of Sciences of Ukraine. 

\vspace*{-.2cm}

\bibliographystyle{elsarticle-num}

\end{document}